\documentclass[conference]{IEEEtran}
\usepackage{cite}
\usepackage{amsmath,amssymb,amsfonts}
\usepackage{algorithmic}
\usepackage{graphicx}
\usepackage{textcomp}
\usepackage{xcolor}
\usepackage{array}
\usepackage{float}
\usepackage{multirow}
\usepackage{subcaption}
\usepackage{tabularx}
\usepackage{wrapfig}
\def\BibTeX{{\rm B\kern-.05em{\sc i\kern-.025em b}\kern-.08em
    T\kern-.1667em\lower.7ex\hbox{E}\kern-.125emX}}
\begin{document}

\title{Investigating the Generalizability of Physiological Characteristics of Anxiety}

\author{\IEEEauthorblockN{Emily Zhou}
\IEEEauthorblockA{\textit{Department of Electrical Engineering} \\
\textit{University of Southern California} \\
Los Angeles, USA \\
emilyzho@usc.edu}
\and
\IEEEauthorblockN{Mohammad Soleymani}
\IEEEauthorblockA{\textit{Institute for Creative Technologies} \\
\textit{University of Southern California} \\
Los Angeles, USA \\
msoleyma@usc.edu}
\and
\IEEEauthorblockN{Maja J. Matarić}
\IEEEauthorblockA{\textit{Department of Computer Science} \\
\textit{University of Southern California} \\
Los Angeles, USA \\
mataric@usc.edu}
}

\maketitle

\begin{abstract}
  \textbf{Recent works have demonstrated the effectiveness of machine learning (ML) techniques in detecting anxiety and stress using physiological signals, but it is unclear whether ML models are learning physiological features specific to stress. To address this ambiguity, we evaluated the generalizability of physiological features that have been shown to be correlated with anxiety and stress to high-arousal emotions. Specifically, we examine features extracted from electrocardiogram (ECG) and electrodermal (EDA) signals from the following three datasets: Anxiety Phases Dataset (APD), Wearable Stress and Affect Detection (WESAD), and the Continuously Annotated Signals of Emotion (CASE) dataset. We aim to understand whether these features are specific to anxiety or general to other high-arousal emotions through a statistical regression analysis, in addition to a within-corpus, cross-corpus, and leave-one-corpus-out cross-validation across instances of stress and arousal. We used the following classifiers: Support Vector Machines, LightGBM, Random Forest, XGBoost, and an ensemble of the aforementioned models. We found that models trained on an arousal dataset perform relatively well on a previously unseen stress dataset, and vice versa. Our experimental results suggest that the evaluated models may be identifying emotional arousal instead of stress. This work is the first cross-corpus evaluation across stress and arousal from ECG and EDA signals, contributing new findings about the generalizability of stress detection.}
\end{abstract}

\maketitle
\pagestyle{empty}

\section{Introduction}
Anxiety and stress are common responses to actual or perceived threats in daily life. While there are negative connotations to anxiety and stress, they are not necessarily harmful to well-being. Acute anxiety and stress may help us to adapt to stressors and respond accordingly, but severe, prolonged stress may give rise to illness. Chronic stress and anxiety can weaken the immune system and have been linked to many health issues, including major neurological disorders, heart diseases, and high blood pressure \cite{Schneiderman_2005}, \cite{Mariotti_2015}. Anxiety disorders and chronic stress are highly prevalent; in the United States alone, with two in three adults report that they have experienced increased stress over the course of the COVID-19 pandemic \cite{AIS}, and rates of moderate to severe anxiety have increased and remain above pre-pandemic levels \cite{MHA_2021}. Due to the closely related emotional experiences of anxiety and stress \cite{stress_vs_anxiety}, we refer to both as "stress" in this paper. 

The growing mental health crisis has inspired an increased interest in developing computing tools to aid the early detection of chronic anxiety and stress. However, to the best of our knowledge, there is less work on generalized methods for anxiety, stress, and affect detection. In this work, we perform cross-corpus stress and emotion recognition using physiological features typically indicative of heightened stress and anxiety across three datasets of stress and general affect in order to examine the generalization properties of these physiological features. 

In general, cross-corpus emotion recognition is a more challenging task due to the different types of stimuli, signal quality, and emotion models used in annotation \cite{Schuller}, \cite{Zehra_2021}, \cite{Zhang_2011}. Studies of the generalization of emotion recognition methods typically classify emotions as high/low arousal or positive/negative valence across different datasets, while generalization of stress detection methods focuses on classifying high/low stress. {\it Few studies have investigated the generalization of stress detection methods, and to the best of our knowledge, none have attempted to evaluate the generalization of emotion and stress detection methods in a true cross-corpus evaluation}, though prior works have performed leave-one-corpus-out evaluations and/or train-test analyses with data pooled from multiple datasets.

Emotions are often described using the axes of arousal and valence \cite{Russell}, which we describe in greater detail in Section \ref{sec:related-work}. Higher stress is indicative of higher arousal, but higher arousal does not necessarily indicate higher stress. In this work, we performed within-corpus, cross-corpus, and leave-one-corpus-out analyses using three affective datasets of stress and arousal to test the following hypothesis: 

\textbf{H1:} Stress detection models may be learning features that are general across emotional arousal instead of specific to stress.

The three datasets we used are APD (stress) \cite{Senaratne}, WESAD (stress) \cite{Schmidt}, and CASE (arousal) \cite{Sharma}. Because stress is one type of high-arousal emotion, we expect an affect detection model trained on arousal labels to perform relatively well on a previously unseen dataset with stress labels. The converse would verify our hypothesis, \textit{i.e.}, the stress detection models are {\it not} specific to stress and are instead learning features generalizable to high arousal. This work makes the following contributions:
\begin{enumerate}
    \item The first cross-corpus evaluation across stress and arousal using manually extracted physiological signals from ECG and EDA signals, contributing new findings about the generalizability of stress detection. 
    \item Regression analysis of the features identified to be indicative of anxiety, stress, and high arousal, comparing their significance and direction of correlation across corpora. 
\end{enumerate}

\section{Related Work}
\label{sec:related-work}
\subsection{Psychophysiology of Emotional Arousal, Anxiety, and Stress}
\begin{figure}[hb]
    \centerline{\includegraphics[width=0.37\textwidth]{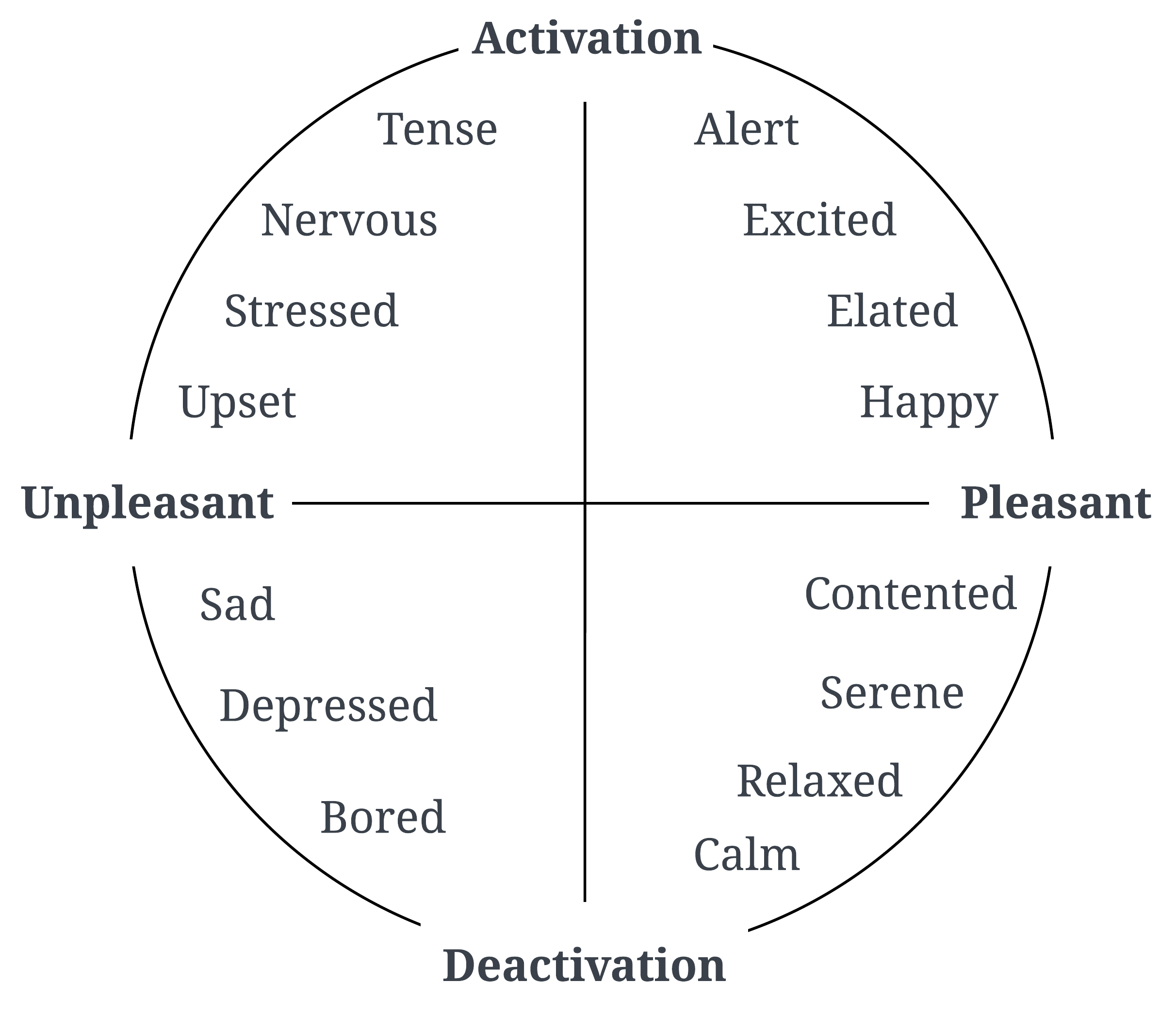}}
    \caption{A graphical representation of the Circumplex Model of affect. The horizontal axis represents the valence dimension and the vertical axis represents the arousal dimension \cite{Posner}.}
    \label{fig:affect-model}
\end{figure}

Among the most commonly used theoretical frameworks for representing emotions is Russell's Circumplex Model of affect \cite{Russell}, which uses the two dimensions of valence (pleasure-displeasure) and arousal (high-low alertness). This model, depicted in Figure \ref{fig:affect-model}, proposes that emotions can be represented on perpendicular axes of valence and arousal \cite{Posner}. Increased emotional arousal is associated with a number of physiological changes caused by autonomic nervous system activity, which include increases in heart rate and the skin conductance response (SCR) amplitude \cite{D'Hondt} and changes in cutaneous blood flow, piloerection, and sweating, measured via electrocardiogram (ECG) and electrodermal activity (EDA). 

Under Russell's Circumplex Model, both anxiety and stress are described as high-arousal emotional states. {\it Anxiety} is defined as a future-oriented mood state of acute sympathetic arousal \cite{Cisler_2010} and is characterized by persistent worries, even in the absence of an external stressor. {\it Stress} is closely related to anxiety but is typically caused by an external trigger that may be short-term or sustained \cite{Alvord_Halfond_2019}. Despite this difference, stress and anxiety exhibit the same physiological features that can be derived from ECG and EDA signals, \textit{e.g.}, increased heart rate, decreased heart rate variability, sympathetic dominance measured using frequency metrics of ECG signals, increased skin conductance levels, and increased skin conductance response rates. We describe each of these in greater detail in Section \ref{sec:data-preprocessing-segmentation}.

The physiological markers of SNS activity related to anxiety, stress, and arousal can be measured easily and cost-effectively with commercially available sensors, and they have been studied extensively by researchers interested in stress detection. Commonly used physiological signals include electrocardiogram (ECG) and electrodermal activity (EDA) \cite{Vos_2023}. Heart rate variability and frequency-domain metrics of ECG signals have been used as biomarkers to measure ANS activity, as well as acute and chronic stress levels \cite{Vos_2023}, \cite{Sepulveda_2021}. Similarly, phasic and tonic (relatively longer-lasting) components of EDA are useful biomarkers for measuring SNS activity and have been used extensively in assessing patients with disorders such as anxiety and depression \cite{Ali_2022}. 

\subsection{Anxiety/Stress Detection Using Physiological Signals}
Many past works have focused on binary or multi-class classification of stress and/or anxiety and emotion classification using physiological signals from publicly available datasets. They primarily focus on developing better-performing models and identifying the most salient feature sets \cite{Bobade}, \cite{Zhu}, \cite{Kuttala}. Previous work on WESAD \cite{Bobade} used statistical features of ECG, EDA, EMG, and temperature data to perform three-class and binary classification, achieving up to 84.32\% and 93.2\% accuracy, respectively, with an artificial neural network. Subsequent studies tested those approaches on multiple datasets; Zhu et al. \cite{Zhu} evaluated a stacking ensemble learning model with EDA, ECG, and PPG signals to examine the accuracy of each modality; they achieved the highest accuracy using only EDA features, up to 86.4\% with WESAD and 72.3\% with the Cognitive Load, Affect, and Stress Database (CLAS). Artificial neural network approaches seeking to achieve better performance than SVM and tree-based methods were also developed \cite{Kuttala}, \cite{Bara}, \cite{Albertetti}. \cite{Kuttala} used frequency-domain features as input to a convolutional neural network (CNN) and tested it on the ASCERTAIN, CLAS, MAUS, and WAUC datasets using low, mid, and high-level features of a CNN. They observed accuracy scores of up to 93.58\% with ASCERTAIN using a multimodal, multi-level feature fusion approach. 

Existing works in stress detection demonstrate the feasibility of using physiological signals from wearable devices to detect stress using SVM, tree-based methods, and artificial neural networks, but they have not determined any ideal combination of modalities or classification techniques. Our approach was guided by previous works that found ML-based classifiers such as SVM, KNN, LightGBM, and XGBoost and an ensemble of those models to be effective due to their high performance and low computational overhead \cite{Zhu}. Previous works found that multimodal feature-level fusion performed better than unimodal features \cite{Kuttala}. Following their success, we use multiple modalities, \textit{i.e.}, high-level and statistical features from ECG and EDA signals, and perform a feature-level fusion. 

\subsection{Cross-Corpus Affect Recognition}
Previous research studying cross-corpus affect recognition primarily focused on general affect recognition using EEG and audiovisual signals. Rayatdoost and Soleymani \cite{Rayatdoost_2018} explored cross-corpus binary emotion recognition using EEG signals and found that model performance decreased when evaluated across corpora. Zehra et al. \cite{Zehra_2021} performed both within- and cross-corpus binary speech emotion recognition (high vs. low valence) using four datasets with different languages. They found that different classifiers performed better on different datasets, and an ensemble learning approach with majority voting generally improved cross-corpus performance. They achieved an F1-score of 0.63 and accuracy of 63.26\% in the cross-corpus setting, training on the Urdu corpus and testing on German. Zhang et al. \cite{Zhang_2011} showed that adding unlabeled speech data to a pooled training set can enhance model performance in both within- and cross-corpus settings. They achieved 66.6\% accuracy in binary affect recognition (high vs. low arousal) using a leave-one-corpus-out validation approach with six different datasets. 

To the best of our knowledge, Baird et al. \cite{Baird_2021} are the only authors who have reported a cross-corpus study using physiological signals for stress recognition. Their analysis focused on predicting sequential saliva-based cortisol measures from the FAU-TSST, ULM-TSST, and REG-TSST datasets. However, the biochemical analysis procedure for cortisol extraction varied between datasets, making the derived values not completely comparable. Rather then training a model on one dataset and testing on another to evaluate the generalizability of their approach, they evaluated a joint regression model in which data were pooled from multiple datasets. They found that respiration and heart rate were highly predictive of cortisol levels, and therefore, effective markers of stress. Furthermore, a review of studies on stress detection using physiological signals and wearable sensors between 2012 and 2022 by \cite{Vos_2023} found no studies attempting to validate a trained model on a completely new, unseen dataset. {\it To the best of our knowledge, we contribute the first study using physiological signals from wearable devices for cross-corpus and cross-label stress and arousal detection.} 

\newcolumntype{b}{X}
\newcolumntype{s}{>{\hsize=.35\hsize}X}

\begin{table}[hb!]
\small
\caption{List of abbreviations used}
\begin{tabularx}{0.45\textwidth} {s|b}
    \hline
    \textbf{Abbreviation} & \textbf{Refers to} \\
    \hline
    ECG & Electrocardiogram \\ 
    \hline
    EDA & Electrodermal activity \\ 
    \hline
    HF-RR & Power in the high-frequency band of RR intervals \\ 
    \hline
    LF-HF ratio & Ratio of LF to HF power \\ 
    \hline
    LF-RR & Power in the low-frequency band of RR intervals \\ 
    \hline
    LGBM & LightGBM \\
    \hline
    RF & Random Forest \\ 
    \hline
    RMSSD & Root mean square of successive differences between normal heartbeats \\ 
    \hline
    SCL & Skin conductance level \\ 
    \hline
    SCR & Skin conductance response  \\ 
    \hline
    SDNN & Standard deviation of the IBI of normal sinus beats \\ 
    \hline
    TSST & Trier Social Stress Test \\ 
    \hline
    XGB & XGBoost \\ 
    \hline
\end{tabularx}
\end{table}

\section{Datasets}
\label{sec:datasets-used}
\subsection{Overview}
We performed stress and affect detection using the Anxiety Phases Dataset (APD), Wearable Stress and Affect Detection (WESAD), and the Continuously Annotated Signals of Emotion (CASE) dataset. In our analysis, we focused on ECG and EDA signals due to their availability across all datasets. Each of these datasets contains data from various wearable devices that collect physiological signals and relies on participant self-reports to generate affective labels. Baseline signals were collected during an initial relaxation phase, and recovery signals were collected during the ending "cool-down" phase in each study. Between each phase of emotion elicitation, subjects were given a short rest period before the next stimulus.

The datasets differ primarily in the method of emotion elicitation and type of self-report used. APD targeted bug phobia and public speaking phobia in two tasks and uses the Subjective Units of Distress scale (SUDs) \cite{McCabe} and the Liebowitz Social Anxiety Scale (LSAS) \cite{Liebowitz_1999}. WESAD used eleven humorous video clips to elicit amusement and the Trier Social Stress Test (TSST) \cite{Kirschbaum_1993}, which consists of a public speaking and a mental arithmetic task, to elicit stress. Self-reports were collected using the Positive and Negative Affect Schedule (PANAS) \cite{Watson_1988}, six-item State-Trait Anxiety Inventory (STAI) scored on a four-point Likert scale \cite{Barker}, Self-Assessment Manikins (SAM) \cite{Bradley_1994}, and Short Stress State Questionnaire (SSSQ) \cite{Helton_2004}. CASE introduced a new method of collecting affect self-reports; participants watched video stimuli and reported their continuous emotional experience using a joystick \cite{Sharma_joystick}, where the x-axis measured valence and the y-axis measured arousal on a scale from 1 to 10. Next, we describe each dataset in greater detail.

\subsection{Anxiety Phases Dataset}
APD consists of electrocardiogram (ECG), electrodermal activity (EDA), and accelerometer (ACC) signals from 52 participants in lab-controlled settings across different phases of induced emotions: rest, anticipation, exposure to an anxiety trigger, recovery, and a spoken reflection of the experience. ECG signals were recorded at 250 Hz, and EDA signals were recorded at 50 Hz. Participants wore a Zephyr BioHarness 3.0 on the chest, and a Grove-GSR Sensor on each wrist. 

Two anxiety triggers were used: a bug-box task in which participants were asked to release a fake bug from a small box, and a speech anxiety task where participants were asked to rank three topics in order of difficulty, then prepare a 3-minute speech about the most difficult topic. At the end of the study, participants reported their anxiety levels using SUDs and LSAS. Each experiment phase lasted 3 minutes.

\subsection{Wearable Stress and Affect Detection}
WESAD is a dataset with stress conditions that includes ECG, EDA, ACC, electromyography (EMG), respiration, and temperature signals collected from 15 subjects across rest, amusement, stress, and two meditation phases. ECG and EDA signals were both recorded at 700 Hz using the chest-worn RespiBAN Professional and wrist-worn Empatica E4. We used signals collected from the chest-worn sensor because the original paper reported slightly higher accuracies using a single chest-worn sensor as opposed to the wrist-worn sensor. Similarly, \cite{Lin_2019} found that chest device features had higher importance than wrist-device features in WESAD.

The TSST was used in the stress condition, and a set of 11 humorous video clips was used in the amusement condition. After each condition, subjects self-reported affect using the PANAS, STAI, SAM, and SSSQ questionnaires. The baseline phase was 20 minutes long, the amusement condition was about 6.5 minutes, the stress condition was 10 minutes, and each meditation phase was 7 minutes.

\subsection{Continuously Annotated Signals of Emotion}
CASE is a multimodal dataset containing data from 30 participants recorded using ECG, blood volume pulse (BVP), galvanic skin response (GSR, which records the same type of signal as EDA), respiration (RSP), skin temperature (TEMP), and electromyography (EMG) sensors. The authors selected 8 videos previously validated in other studies to elicit amusement, boredom, relaxation, and fear, as well as three additional videos aimed at eliciting a state of relaxation for baseline measurements. Each video lasted 2 to 3 minutes.

\section{Methodology}
\subsection{Data Preprocessing, Segmentation, and Feature Extraction}
\label{sec:data-preprocessing-segmentation}
Our preprocessing methods were modeled after those described in \cite{Schmidt} and \cite{Senaratne}. ECG and EDA signals were first filtered using preprocessing methods from the biosppy \cite{biosppy} and NeuroKit \cite{NeuroKit} Python open-source libraries to reduce noise and remove baseline wander. Then, as recommended by \cite{Fang_2022}, for each phase in the datasets, we used 60-second sliding windows with an overlap of 30 seconds to extract the following ECG and EDA metrics from segments of the denoised signal.

\noindent \textbf{Electrocardiogram (ECG):}
\begin{itemize}
    \item BPM: Heart rate increases in the face of a stressor \cite{Noteboom_2001}. BPM was calculated using the biosppy.ecg module.
    \item RMSSD and SDNN: Decreased HRV is associated with elevated anxiety \cite{Kreibig_2010}. RMSSD reflects the beat-to-beat variance in HR and is the primary time-domain measure used to estimate the vagally mediated changes reflected in HRV \cite{Shaffer_2017}. Lower SDNN and RMSSD was reported in subjects with various anxiety disorders in \cite{Licht_2009}, \cite{Prasko_2011}, and \cite{Chalmers_2014}. RMSSD and SDNN were calculated using pyhrv \cite{Gomes_2019}.
    \item HF-RR and LF-RR: The high-frequency band of HR oscillations ranges from 0.15–0.4 Hz, and the low-frequency band of HR oscillations ranges from 0.04-0.15 Hz \cite{Shaffer_2017}. The HF band reflects parasympathetic influences, and lower HF power is correlated with stress, panic, anxiety, or worry. Increased LF power is associated with anxiety \cite{Prasko_2011}. In general, LF power may be generated by the sympathetic nervous system \cite{Shaffer_2017}. The Fast Fourier Transform was used to calculate HF-RR and LF-RR.
    \item LF-HF ratio: Higher LF/HF ratio indicates sympathetic dominance, which occurs during fight-or-flight responses or decreased parasympathetic activity \cite{Shaffer_2017}. In healthy individuals, acute stress increases the LF/HF ratio and decreases HF \cite{Pagani_1997}; \cite{Kreibig_2010} describes other studies that have reported the same pattern.
\end{itemize}

\noindent \textbf{Electrodermal Activity (EDA):}
\begin{itemize}
    \item Mean SCL: The skin conductance level is the tonic component of EDA. Changes in the SCL are thought to reflect general changes in autonomic arousal \cite{Braithwaite_2015}, and increased SCL is associated with anxiety \cite{Ali_2022}. SCL was calculated using the NeuroKit Python library, then averaged over the 60-second window. 
    \item SCR rate: Skin conductance response is the phasic component of EDA \cite{Ali_2022} and a higher skin conductance response rate has been associated with anxiety, threat responses, and emotional arousal \cite{Craske_2022}. SCR rate was also calculated using NeuroKit.
\end{itemize}

We also extracted the mean, median, standard deviation, and variance from filtered ECG and EDA signals. 

\subsection{Label Generation}
Each dataset used a different method to collect self-reports, so it was necessary to process those data in a way that allowed for cross-corpus comparison.

\noindent \textbf{APD:}
Participants’ self-reported anxiety levels from the SUDs questionnaire were used to generate binary labels. The version of SUDs used in APD ranges from 0 to 100, and we used 50 as a fixed threshold of the median score, to differentiate high from low anxiety. Participant phases with scores 50 and above were labeled as 1 (high anxiety) and those below 50 were labeled as 0 (low anxiety). 

\noindent \textbf{WESAD:}
Participants’ STAI responses were used to generate binary labels. The 6-item short form STAI questionnaire \cite{short_STAI} ranges from 6 to 24, and we used the median, 15, as the fixed threshold to determine high vs. low anxiety. 

\noindent \textbf{CASE:}
Participants’ self-reports of arousal were used to generate binary labels. Since CASE consists of continuous self-reports, we averaged the self-reported arousal values for each phase to be consistent with APD and CASE. The median value of the joystick range, 5, was used as the fixed threshold. 

\subsection{Statistical Analysis and Model Training}
We first used a linear mixed effects model from the statsmodels Python library \cite{statsmodels} to examine the effects of the selected physiological features on stress and emotional arousal within each dataset. We set the physiological features as fixed effects and the subject group as random effects. 

Next, we conducted experiments in within-corpus, cross-corpus, and leave-one-corpus-out settings. Our results from the within-corpus condition are used as baselines for the cross-corpus and LOCO experiments. Five machine learning techniques were used to perform binary classification: Support Vector Machine (SVM), LightGBM (LGBM) \cite{LGBM}, Random Forest (RF), XGBoost (XGB), and an ensemble of the previous four models. These models were chosen based on the SOTA performance achieved by SVM, KNN, and tree-based classifiers in previous affect detection studies, discussed in Section \ref{sec:related-work}. We evaluated both majority voting and equally weighted average ensembles of models. The equally weighted ensemble approach performed significantly better, so we omit the majority voting results. 

In each experimental setting, data samples were generated from the extracted features by segmenting each phase into 2-minute chunks and calculating the mean for each chunk. Rather than take the mean value over the whole phase, we chose 2-minute segments, since the shortest phase across all three datasets was 2 minutes long. This process was repeated for each feature and phase for each participant. Then, the features for each phase were concatenated to form the input feature vector, which was then normalized before model training. In cases of moderate class imbalance, SMOTE \cite{SMOTE} was used. Model hyperparameters were selected using a simple grid search algorithm. 

In the cross-corpus and leave-one-corpus-out (LOCO) experiments, we used the following definitions of binary classification: when testing on APD or WESAD, it is defined as classifying an individual as having either high or low anxiety, and when testing on CASE, it is defined as classifying an individual as having either high or low emotional arousal, regardless of the labels in the training dataset used.

\begin{table*}[]
\centering
\caption{Correlation coefficients, \textit{p}-values, and \textit{R}-squared values (conditional/marginal) obtained from a mixed linear model fitted to each dataset. An asterisk (*) indicates significance, i.e., a two-tailed p-value of less than 0.1, or equivalently, a one-tailed p-value of less than 0.05.}
\label{tab:reg-analysis}
\begin{tabular}{l|cc|cc|cc}
            & \multicolumn{2}{c|}{APD}                    & \multicolumn{2}{c|}{WESAD}                  & \multicolumn{2}{c}{CASE}                   \\ \hline
            & \multicolumn{2}{l|}{R-squared: 0.688/0.120} & \multicolumn{2}{l|}{R-squared: 0.797/0.556} & \multicolumn{2}{l}{R-squared: 0.629/0.325} \\ \hline
            & Coef      & p-value      & Coef      & p-value      & Coef      & p-value      \\ \hline
BPM         & 0.239     & 2.32E-12*    & 0.441     & 1.01E-79*    & 0.254     & 2.47E-07*    \\
RMSSD       & -0.013    & 0.771        & 0.17      & 1.54E-13*    & 0.087     & 0.027*       \\
HF\_RR      & 1.048     & 0.360        & 0.053     & 0.486        & 0.645     & 0.007*       \\
LF\_RR      & -1.286    & 0.268        & 0.033     & 0.689        & -0.769    & 0.002*       \\
SDNN        & -0.015    & 0.713        & -0.105    & 3.18E-06*    & -0.032    & 0.339        \\
Mean SCL    & 1.74      & 0.338        & -0.037    & 0.043*       & 0.349     & 9.90E-13*    \\
SCR rate    & 0.204     & 0.1*         & -0.066    & 3.55E-06*    & -0.034    & 0.304        \\
LF/HF ratio & 0.055     & 0.153        & 0.121     & 3.20E-06*    & 0.049     & 9.64E-08*    \\
ECG mean    & -1.275    & 0.037*       & -0.148    & 0.558        & -1.694    & 0.144        \\
ECG median  & 1.295     & 0.037*       & 0.134     & 0.601        & 1.653     & 0.154        \\
ECG std     & -0.075    & 0.306        & 0.038     & 0.688        & 1.024     & 2.56E-24*    \\
ECG var     & -0.041    & 0.559        & -0.157    & 0.07*        & -0.936    & 3.44E-21*    \\
EDA mean    & -1.82     & 0.318        & -0.03     & 0.007*       & -0.594    & 0.001*       \\
EDA median  & 0.019     & 0.894        & 0.063     & 0.007*       & 0.234     & 0.162        \\
EDA std     & 0.058     & 0.246        & 0.065     & 0.06*        & -0.094    & 0.107        \\
EDA var     & -0.041    & 0.406        & -0.024    & 0.419        & 0.134     & 0.017*                     
\end{tabular}
\end{table*}
\begin{table}[tbp!]
\caption{Within-corpus classification results for stress vs. non-stress (Acc and AUC score) in APD and WESAD and high vs low arousal in CASE. We use SVM, LightGBM, Random Forest, XGBoost, and an ensemble of the previous models. The best results across all classifiers are bolded.}
\label{tab:within-corpus}
\begin{tabular}{c|cc|cc|cc}
\multicolumn{1}{c|}{Model}    & \multicolumn{1}{c}{Acc}        & \multicolumn{1}{c|}{AUC}       & \multicolumn{1}{c}{Acc}         & \multicolumn{1}{c|}{AUC}        & \multicolumn{1}{c}{Acc}        & \multicolumn{1}{c}{AUC}       \\ \hline
         & \multicolumn{2}{c|}{APD} & \multicolumn{2}{c|}{WESAD} & \multicolumn{2}{c}{CASE} \\ \hline
Random   & 0.471           & 0.467           & 0.509            & 0.432            & 0.471           & 0.503           \\\hline
SVM      & 0.549           & 0.498           & 0.860            & 0.720            & 0.800           & 0.658           \\
LGBM     & 0.547           & 0.529           & 0.836            & 0.720            & 0.788           & 0.661           \\
RF       & 0.553           & 0.540           & 0.845            & 0.735            & 0.792           & 0.692           \\
XGB      & 0.561           & 0.528           & 0.842            & 0.775            & 0.767           & 0.671           \\\hline
Ensemble & \textbf{0.645}  & \textbf{0.608}  & \textbf{0.990}   & \textbf{0.969}   & \textbf{0.813}  & \textbf{0.736} 
\end{tabular}
\end{table} 

\noindent \textbf{Within-corpus: }
In the within-corpus case, models were first trained and tested within datasets, using a 5-fold cross-validation technique. The participants were split into 5 equally sized groups, and data instances from each of the phases were aggregated to form each fold.  

\noindent \textbf{Cross-corpus: }
In the cross-corpus setting, we trained and tested classifiers across pairs of datasets. One dataset was used as the training set, while the other was held out as the test set to evaluate the generalizability of the features and models. 

\noindent \textbf{Leave-one-corpus-out: }
Finally, we conducted a LOCO experiment using SVM, LGBM, RF, XGB, and an ensemble of the four classifiers. Two datasets were pooled and used as input to the model, while the remaining dataset was held out as the test set. 

\section{Results and Discussion}
\label{sec:results-discussion}
\subsection{Regression Analysis}
\label{sec:regression-analysis}
Results from the regression analysis are reported in Table \ref{tab:reg-analysis}. Heart rate (BPM) was significant across all three datasets, while SCR rate was significant and positively correlated in both APD and WESAD. We found that half of the features had the same direction of correlation across all three datasets. BPM, HF-RR, LF-RR, the median of ECG, and the median of EDA were positively correlated to a positive label, and SDNN, the mean and variance of ECG, and the mean of EDA were negatively correlated. Several of these correlations are consistent with previous work in psychology and affect detection: increased heart rate \cite{Noteboom_2001} and power in the low-frequency band \cite{Prasko_2011} have been identified as high-level features indicative of SNS activity. SDNN is a measure of HRV, and decreased HRV is known to be correlated with increased stress \cite{Kreibig_2010}. While SCR rate has been found to be positively correlated with stress and anxiety \cite{Craske_2022}, we find a slightly positive correlation in WESAD and CASE. This could be due to the elicitation of amusement in both datasets, which was found to have a lower skin conductance than fear in \cite{Leng_2007}. \textit{Given that the majority of physiological features exhibited the same direction of correlation in all three datasets, the statistical analysis results support the generalizability of these features across stress and anxiety.}

\begin{table}[tbp!]
\caption{Cross-corpus classification results using SVM, LightGBM, Random Forest, XGBoost, and an ensemble of the previous models (accuracy and AUC score). Random classifier performance is included as a point of comparison.}
\label{tab:cross-corpus}
\begin{tabular}{c|cc|cc|cc}
Model    & \multicolumn{1}{c}{Acc}       & \multicolumn{1}{c|}{AUC}      & \multicolumn{1}{c}{Acc}       & \multicolumn{1}{c|}{AUC}      & \multicolumn{1}{c}{Acc}       & \multicolumn{1}{c}{AUC}      \\ \hline
& \multicolumn{2}{c|}{APD/WESAD}    & \multicolumn{2}{c|}{APD/CASE}     & \multicolumn{2}{c}{WESAD/CASE}    \\ \hline
Random   & 0.509          & 0.432          & 0.471          & 0.503          & 0.471          & 0.503          \\ \hline
SVM      & 0.358          & 0.530          & 0.458          & 0.500          & 0.463          & 0.510          \\
LGBM     & 0.434          & 0.597          & 0.610          & 0.611          & \textbf{0.696}          & \textbf{0.617}          \\
RF       & 0.260          & 0.515          & 0.515          & 0.564          & 0.598          & 0.582          \\
XGB      & 0.499          & 0.641          & 0.569          & \textbf{0.644}          & 0.610          & 0.626          \\
Ensemble & \textbf{0.658}          & \textbf{0.709}          & \textbf{0.708}          & 0.611          & 0.644          & 0.600          \\ \hline
& \multicolumn{2}{c|}{WESAD/APD}    & \multicolumn{2}{c|}{CASE/APD}     & \multicolumn{2}{c}{CASE/WESAD}    \\ \hline
Random   & 0.471          & 0.467          & 0.471          & 0.503          & 0.509          & 0.432          \\ \hline
SVM      & 0.552          & 0.539          & 0.535          & 0.522          & \textbf{0.774}          & \textbf{0.649}          \\
LGBM     & 0.550          & 0.542          & 0.546          & 0.537          & 0.680          & 0.555          \\
RF       & 0.513          & 0.527          & 0.538          & 0.544          & 0.559          & 0.637          \\
XGB      & 0.503          & 0.550          & 0.536          & 0.529          & 0.755          & 0.605          \\ \hline
Ensemble & \textbf{0.569}          & \textbf{0.561}          & \textbf{0.567}          & \textbf{0.551}          & 0.759          & 0.591
\end{tabular}
\end{table}
\begin{table}[tbp!]
\caption{Leave-one-corpus out classification results using SVM, LightGBM, Random Forest, XGBoost, and an ensemble of the previous models (accuracy and AUC score).}
\label{tab:loco}
\begin{tabular}{c|cc|cc|cc}
\multicolumn{1}{c|}{Model}        & \multicolumn{1}{c}{Acc}    & \multicolumn{1}{c|}{AUC}    & \multicolumn{1}{c}{Acc}         & \multicolumn{1}{c|}{AUC}   & \multicolumn{1}{c}{Acc}       & \multicolumn{1}{c}{AUC}      \\ \hline
& \multicolumn{2}{c|}{Test: APD} & \multicolumn{2}{c|}{Test: WESAD} & \multicolumn{2}{c}{Test: CASE}  \\ \hline
Random    & 0.471          & 0.467          & 0.509          & 0.432          & 0.471          & 0.503 \\ \hline
SVM       & 0.501          & 0.492          & 0.707          & 0.510          & 0.621          & 0.546 \\
LGBM      & 0.505          & 0.497          & 0.590          & 0.475          & 0.535          & 0.534 \\
RF        & 0.515          & \textbf{0.513}          & \textbf{0.753}          & 0.536          & 0.463          & \textbf{0.562} \\
XGB       & 0.516          & 0.507          & 0.182          & \textbf{0.571}          & 0.492          & 0.541 \\ \hline
Ensemble  & \textbf{0.521}          & 0.506          & 0.712          & 0.506          & \textbf{0.610}          & 0.552
\end{tabular}
\end{table}

\subsection{Within-Corpus Performance}
\label{sec:results-within-corpus}
We use accuracy and AUC metrics to analyze model performance in the within-corpus, cross-corpus, and LOCO settings. Table \ref{tab:within-corpus} presents our results in the within-dataset condition, as well as the results achieved with a random classifier to serve as a baseline for the within-dataset experiment. All models performed significantly better than a random classifier, and we achieve SOTA performance on WESAD in the within-dataset setting compared to previous binary classification studies \cite{Bobade}, \cite{Zhu} using an ensemble of models. The ensemble of models performed the best overall across all datasets. It is important to note that our goal in the within-dataset case was not to achieve SOTA performance across all datasets. This step verifies that our pipeline, from preprocessing to model training and classification, is a viable method of affect detection for the APD, WESAD, and CASE datasets. Model performance is the lowest in APD, which may be attributable to more noise present in the data. These results verify that our preprocessing, feature extraction, and classification methods are viable for detecting stress and arousal.

\subsection{Specificity of Stress Detection: Cross-Corpus and Leave-One-Corpus-Out Performance}
Cross-corpus results are presented in Table \ref{tab:cross-corpus}, and LOCO results in Table \ref{tab:loco}. We found that model performance in cross-corpus stress and affect detection was lower than within-corpus performance, consistent with the findings in \cite{Rayatdoost_2018}. However, we still achieved better-than-chance accuracy and AUC scores across all pairs of datasets.

Our goal was to identify whether stress detection models and the selected physiological features are specific to stress or generalizable to arousal using cross-corpus analyses across three datasets. The cross-corpus and LOCO experiments were designed to evaluate the specificity of stress detection by testing models with unseen data and different labels (\textit{i.e.}, stress vs arousal) from a completely different distribution. In the case where both training and testing datasets were labeled with stress (APD/WESAD), we achieved up to 68.3\% accuracy and an AUC score of 0.761 testing on WESAD, and 56.9\% accuracy and an AUC score of 0.561 testing on APD.  Some train-test instances were not much better than random guessing, \textit{e.g.} testing on APD regardless of the training dataset in both cross-corpus and LOCO cases. This is unsurprising given the lower performance for APD in the within-corpus case; the discrepancy may be due to a higher signal-to-noises ratio present in APD data.

To evaluate our hypothesis, we examine the cross-label training results, \textit{i.e.}, training on a stress dataset and testing on an arousal dataset, and vice versa. Since stress is an example of a high arousal emotion, we expected models trained on arousal labels to perform relatively well on new instances with stress labels, and models trained on stress labels to be specific to stress and therefore perform worse on a dataset labeled with arousal. Instead, we found that training on APD and testing on CASE resulted in better performance (accuracies up to 70.8\% and AUC scores up to 0.644) than training on CASE and testing on APD (accuracies up to 56.7\% and AUC scores up to 0.551). Training on WESAD and testing on CASE resulted in accuracies up to 69.6\% and AUC scores up to 0.617. CASE/APD achieved a maximum accuracy of 77.4\% and AUC score of 0.649.

Therefore, our results are supportive of H1: {\it emotion classification models trained on stress and tested on arousal labels can perform nearly as well or better than training on arousal and testing on stress. In fact, cross-label evaluations resulted in higher accuracies and AUC scores than training and testing between stress datasets.} It appears that the stress detection models we tested may be learning features \textit{general across emotional arousal} instead of \textit{specific to stress}. While our early work in the specificity of stress features is limited in that we did not perform explainable AI techniques such as the calculation of Shapley values, our results suggest that this is an important direction for future work in stress detection. 

\section{Conclusion}
Our work contributes the first study of cross-corpus stress and affect detection across datasets labeled with stress and emotional arousal. We extracted physiological features from ECG and EDA signals and performed stress and arousal detection in within-corpus, cross-corpus, and leave-one-corpus-out experiments on three affective datasets. Our results show that stress recognition models are able to recognize emotional arousal better than stress in difficult cross-corpus settings with different emotional stimuli, suggesting that such recognition models may be detecting an emotionally aroused state that is more general than stress. A limitation of this work is the relatively small number of datasets and features used. We aimed to use common high-level features of ECG and EDA signals that have been found to be indicative of ANS activity, but deep learning-based features are also highly useful in detecting stress and anxiety, which can be explored in future cross-corpus work. We also did not identify a set of physiological features that are specific to stress/anxiety, which merits further investigation with metrics and tools such as Shapley values and feature importances. However, our results suggest the importance of cross-corpus and cross-label experiments to verify that features learned by stress detection model are specific to stress. 

This work motivates a deeper evaluation of stress detection and the development of models that use features specific to stress to accurately identify only stressful instances. To support the improvement of stress detection models toward in-the-wild deployments, more work is needed in cross-corpus settings using larger datasets and more high-arousal emotions. Future work may benefit from collecting physiological signals from a wider variety of stressors and in-the-wild environments to further explore the generalizability of stress detection across settings.

\section*{Acknowledgment}
   This work was supported in part by the National Science Foundation under Grant No. 2211550 and the University of California's Center for Undergraduate Research in Viterbi Engineering (CURVE) fellowship. The authors thank Prof. Bhaskar Krishnamachari for his insights and expertise throughout the development and exploratory analyses of this work.

\bibliographystyle{IEEEtran}
\bibliography{references}
\end{document}